# Quantum State Engineering with the *rf*-SQUID

Christopher Altman[i]

Pierre Laclede Honors College / University of Amsterdam

Home: http://www.umsl.edu/~altmanc/
Contact: contact@artilect.org

## Introduction

The SQUID, or superconducting quantum interference device, is a highly sensitive instrument employed for non-destructive measurement of magnetic fields, with a host of applications in both biophysics and materials technology. It is composed of a cooled superconductive metal ring separated by a thin insulating barrier of non-superconducting metal, forming a Josephson junction. Electron tunnelling through the junction can allow for the measurement of electromagnetic field fluctuations as minute as $10^{-15}$ Tesla, or one femto-Tesla – some $10^{-11}$ times less than the earth's natural magnetic field. An *rf*-SQUID is essentially a Josephson junction with tunable current and energy.

## Quantum Computing

Quantum computers take advantage of the superpositional logic of quantum mechanics to allow for dramatic increases in computational efficiency. *rf*-SQUIDs show potential for quantum computing applications by forming the qubit component of a quantum computer, through simply treating the direction of the persistent current – clockwise or countercockwise – as the value of the bit.

---

[i] Chairman, First Committee on Disarmament and International Security, UNISCA

*rf*-SQUIDs present a major advantage over atomic-scale qubit systems: they are sensitive to parameters that can be engineered. Flux qubits can be linked through controlled inductive coupling - the magnetic field of each junction affects the others. The strength of this coupling can be 'tuned,' allowing for refined control over the behaviour of the system. *rf*-SQUIDs can also be mass produced on-chip, making large-scale production a feasible endeavor.

## Algorithm Design

Algorithm design for the quantum computer poses a number of challenges including initialization, error correction and decoherence time. The unitary transformation represents the most common algorithm performed upon qubits to produce the desired solution from the set of all possible solutions: qubit arrays are manipulated under matrix multiplication to derive the correct solution upon decoherence. An *rf*-SQUID qubit biased at one-half flux is set to a distinct state, the unitary algorithm is performed, and its its state is projected by classical measurement. Unitary matrix transformations are reversible, requiring zero energy to change states or gate operations. Kak outlines:

> The Power of Quantum Computing
>
> We now take a brief look at the question of harnessing the power of quantum computing for the design of AI machines. The dynamics of an isolated quantum system are governed by the Schrodinger equation which can be cast in a form where the future states of the system are obtained by multiplication by a unitary matrix, $U$, whose conjugate transpose is equal to its inverse:
>
> $$| S_{t+1} = U | S_t \rangle$$
>
> The task of the algorithm designer is to first find the unitary matrix for the given computing problem and then map the matrix into a sequential product of smaller matrix operations that can be implemented relatively easily. Effectively, a quantum computation is nothing more than matrix multiplication.
>
> A basic issue in quantum computing is to separate the 'good' solution from the many other data sequences that are simultaneously present on the quantum register, and this must be done without 'looking', because interaction with the contents of the register will cause the superposition state to collapse into one of its components. This separation is achieved by strengthening the amplitude of the desired state by changing the difference in the phase angles of the marked and the unmarked states.

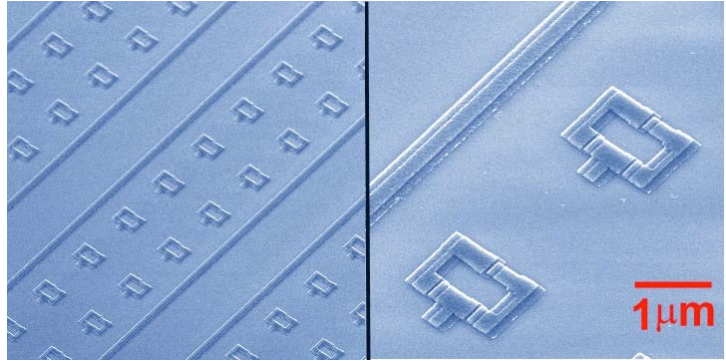

Qubits surrounded by a SQUID line [ Technical University of Delft ]

## Applications

Quantum computing shows great potential for solving problems traditionally considered computationally intractable. The protein folding problem illustrates an inherent weakness in von Neumann computation: our computers cannot effectively simulate the three-dimensional conformations involved in tertiary amino acid folding sequences. This is why we cannot quickly derive the meaning of the complete human DNA sequence, despite our possession of its syntactical mapping. The building blocks are simple, but the emergent structures are vastly complex. This complexity barrier poses the greatest obstacle in the development of disease treatments.

Quantum computing's 'power applications' lie in problems like these which have such a large search space. Kak writes

> It has been estimated that a fast computer applying plausible rules for protein folding would need $10^{127}$ years to find the final folded form for even a very short sequence of just 100 amino acids. Such a mathematical formulation of the protein-folding problem shows that it is NP-complete. Yet Nature solves this problem in a few seconds. Since quantum computing can be exponentially faster than conventional computing, it could very well be the explanation for Nature's speed. The anomalous efficiency of other biological optimisation processes may provide indirect evidence of underlying quantum processing if no classical explanation is forthcoming.

## Challenges

The creation of a functional *rf*-SQUID based quantum computer remains a long-term objective – the path between feasability and implementation remains largely unmapped. Josephson junctions require temperatures on order of degrees milli-Kelvin, and they require near-perfect physical isolation to prevent decoherence. This last step is critical in making the *rf*-SQUID viable for quantum computing applications – the macroscopic size of *rf*-SQUID based qubit arrays makes them uniquely susceptible to environmental disturbance. Better isolation, measurement and error correction techniques will be required to link *rf*-SQUID based qubits together for quantum computation.

Future designs of the *rf*-SQUID - based quantum computer will require a firm understanding of multi-qubit interaction in mesoscopic systems. Bose-Einstein Condensates may present a new opportunity to examine this interaction: recently, Italian researchers created a Bose-Einstein condensate using a magnetic trap and a laser standing wave, producing a Josephson current in the energy barrier between quantum wells. These one-dimensional Josephson arrays, previously difficult to study, present a new avenue for exploration of the behavioural dynamics of the phenomena.